\documentclass[a4paper,11pt]{article}
\pdfoutput=1

\usepackage{jheppub}
\usepackage[T1]{fontenc}
\usepackage{lmodern}
\usepackage{mathtools}
\usepackage{microtype}

\newcommand{\R}{\mathbb{R}}
\newcommand{\C}{\mathbb{C}}
\newcommand{\cH}{\mathcal{H}}
\newcommand{\cD}{\mathcal{D}}

\newcommand{\dd}{\mathrm{d}}
\newcommand{\ii}{\mathrm{i}}
\DeclareMathOperator{\diver}{div}
\DeclareMathOperator{\cof}{cof}

\title{\boldmath Non-injective field redefinitions and quantum inequivalence in scalar theories}

\author{Bin Zhu}
\affiliation{School of Physics, Nankai University,\\
Weijin Road 94, Tianjin 300071, P.R. China}
\emailAdd{bzhu@nankai.edu.cn}

\abstract{We study scalar theories obtained by pulling a free massive
multiplet back through a polynomial field redefinition with constant unit
Jacobian.  Our main example uses the three-variable noninjective map recently
announced by Alp\"oge.  After a linear normalization, it defines a
three-scalar sigma model with a flat, unit-volume field-space metric and three
isolated vacua.  Each vacuum is locally described by three free modes of mass
\(m\), and the exact equations of motion reduce locally on each sheet to free
Klein--Gordon equations.  The global theory is nevertheless not a single
free theory: the field-space metric is incomplete, the number of real
preimages changes across target space, and the commuting position operators
have nonconstant joint spectral multiplicity.  This rules out a global
unitary implementation of the field redefinition and a regular Weyl
exponentiation of the formal canonical momenta.  We then analyze a
four-scalar map with a generic quintic fiber.  It exhibits the same mechanism
with an additional field that controls the fiber polynomial.  The two
examples separate perturbative equivalence on a chosen local sheet from
global quantum equivalence of the full field space.}

\keywords{Field Redefinitions, Canonical Quantization, Sigma Models,
Vacuum Structure}

\begin{document}
\maketitle
\flushbottom

\section{Introduction}
\label{sec:introduction}

Field redefinitions are normally used as local changes of coordinates on
configuration space.  When the transformation is invertible, the
equivalence theorem relates the on-shell descriptions obtained before and
after the change of variables
\cite{Chisholm1961,Kamefuchi1961}.  Modern formulations make the assumptions
behind this statement explicit, including perturbative invertibility,
boundary conditions, and the choice of observables
\cite{CriadoPerezVictoria2019,CohenLuSutherland2024,CriadoJaeckelSpannowsky2025}.
The present work asks what remains true when the Jacobian is nonsingular
everywhere but the transformation is not globally one-to-one.

The closest field-theory precedent is
ref.~\cite{CriadoJaeckelSpannowsky2025}, which explains how a non-one-to-one
field redefinition can change the field-space domain and the set of classical
solutions.  Its representative examples pass through singular loci of the
field map.  The models studied here instead have a nonsingular Jacobian at
every finite field value.  The obstruction therefore comes from the
multiplicity of real preimages rather than a local singularity; no global
inverse is implied.  The Jacobian conjecture has
also been formulated in perturbative and combinatorial quantum field theory
\cite{Abdesselam2003,Tanasa2021}.  Those works use field-theory methods to
organize formal inverse problems; here a real noninjective map is used as the
field-space transformation of a scalar theory.

This question can be studied without introducing a local Jacobian anomaly.
Let
\begin{equation}
  \mathcal F:\R^n\longrightarrow\R^n,
  \qquad
  \det J_{\mathcal F}=1,
  \label{eq:intro-map}
\end{equation}
be a polynomial local diffeomorphism for which some target values have more
than one real preimage.  Throughout this paper, an inverse branch means a
local inverse
\(\mathcal G_s:\mathcal V\to\mathcal U_s\) supplied by the inverse function
theorem on a specified target patch \(\mathcal V\).  No global single-valued
inverse of \(\mathcal F\) exists, and no such object is used below.  Its
nonexistence is the source of the global effect studied here.
Pulling a free massive multiplet back through \(\mathcal F\) produces a
nonlinear sigma model.  Its field-space metric is locally Euclidean and its
volume density is one.  The derivative and potential interactions are
therefore removable around any chosen vacuum.  They are not removable by one
global field coordinate if different points of the original field space are
mapped to the same free-field configuration.

Our simplest realization starts from the three-variable map announced by
Alp\"oge \cite{Alpoge2026Map}.  The announcement is an algebraic result: it
gives a polynomial map with constant Jacobian and an explicit three-point
fiber.  It does not specify a scalar action, particle interpretation, or
equations of motion.  We supply that physics construction here.  A linear
normalization gives a unit-Jacobian map \(\mathcal F_3\), and the pullback of
three free massive scalars has three finite vacua.  The equations of motion
are exactly free in the local variables
\(\mathcal Q^I=\mathcal F_3^I(\varphi)\), but the global Hilbert-space
representation retains the multiplicity of the different preimages.

The quantum distinction is already visible in the commuting position
operators.  On one open target region their joint spectral multiplicity is
three, while on another it is one.  A standard Schr\"odinger coordinate
tuple has multiplicity one, so no unitary operator can implement the
noninjective redefinition globally.  The same variation prevents the formal
canonical momenta from exponentiating to a regular representation of the
Weyl relations.  This is a global obstruction, not a failure of the local
commutators.

We then turn to a four-scalar map whose fibers are governed by a quintic
polynomial.  The fourth field is not a passive coordinate: it enters the first
two output polynomials and controls the coefficients of the fiber equation.
The resulting model has the same local free-field description and the same
global multiplicity obstruction, but a different algebraic realization.  A
generic-degree-five three-scalar slice of this family is useful for comparison
and is retained in
appendix~\ref{app:degree-five-slice}; it is not used as the primary example.

The paper is organized as follows.  Section~\ref{sec:general-framework}
sets up the pullback action and the global quantum diagnostic.
Section~\ref{sec:alpoge-three-scalar} constructs the three-scalar model from
the Alp\"oge map.  Section~\ref{sec:four-scalar} develops the four-scalar
example.  Section~\ref{sec:comparison} compares the two mechanisms, and
section~\ref{sec:discussion} discusses their physical interpretation.  The
generic-degree-five three-scalar slice is retained in the appendix as a
useful intermediate example.

\section{Non-injective field redefinitions}
\label{sec:general-framework}

\subsection{Pulling back a free massive multiplet}

We work first on a fixed Euclidean background \(h_{\mu\nu}\).  Let
\(\varphi^a(X)\), \(a=1,\ldots,n\), be dimensionless real scalar fields and
let \(\mathcal F^I(\varphi)\) be a polynomial map satisfying
eq.~\eqref{eq:intro-map}.  The target variables
\begin{equation}
  \mathcal Q^I(X)=\mathcal F^I(\varphi(X))
  \label{eq:general-target-fields}
\end{equation}
are good local coordinates at every finite field value.  For a fixed target
\(\mathcal Q_\star\), we consider
\begin{equation}
\begin{split}
  S_n[\varphi]
  =\frac{f_\phi^2}{2}\int\dd^dX\,\sqrt h\,
  \big[
    h^{\mu\nu}g_{ab}(\varphi)
    \partial_\mu\varphi^a\partial_\nu\varphi^b
    +m^2\lvert\mathcal F(\varphi)-\mathcal Q_\star\rvert^2
  \big],
  \label{eq:general-action}
\end{split}
\end{equation}
with
\begin{equation}
  g_{ab}(\varphi)
  =\partial_a\mathcal F^I\,\partial_b\mathcal F^I
  =(J_{\mathcal F}^{\mathsf T}J_{\mathcal F})_{ab}.
  \label{eq:general-metric}
\end{equation}
Here \(m\) is the physical mass and \(f_\phi\) sets the field-space scale.
In \(d\) spacetime dimensions,
\([f_\phi]=(d-2)/2\) and \([m]=1\).  The canonically dimensioned local
fluctuations are
\(q^I=f_\phi(\mathcal Q^I-\mathcal Q_\star^I)\).
All integer and rational coefficients inside \(\mathcal F\) are fixed
dimensionless interaction coefficients.  Once the fields are canonically
normalized, nonlinear vertices are organized by inverse powers of
\(f_\phi\).

No spacetime curvature coupling is required for the construction.
Equation~\eqref{eq:general-action} is already minimally covariant on a fixed
curved background.  We set \(h_{\mu\nu}=\delta_{\mu\nu}\) in the explicit
calculations below.  Additional terms such as
\(\mathcal R\lvert\mathcal F-\mathcal Q_\star\rvert^2\) may be added, but
they define a different model and are not needed for the global effect.

On a target patch \(\mathcal V\) over which a local inverse
\(\mathcal G_s:\mathcal V\to\mathcal U_s\) has been chosen,
\(\mathcal Q=\mathcal F(\varphi)\) is a valid local field coordinate.  On the
corresponding source patch \(\mathcal U_s\), the action becomes
\begin{equation}
  S_n
  =\frac{f_\phi^2}{2}\int\dd^dX\,\sqrt h\,
  \left[
    h^{\mu\nu}\partial_\mu\mathcal Q^I\partial_\nu\mathcal Q^I
    +m^2(\mathcal Q^I-\mathcal Q_\star^I)^2
  \right].
  \label{eq:general-free-action}
\end{equation}
Thus the complicated derivative and potential interactions in the
\(\varphi\) variables are redundant in perturbation theory around a fixed
local sheet.  The metric is positive and flat, and
\begin{equation}
  \det g=(\det J_{\mathcal F})^2=1.
  \label{eq:general-detg}
\end{equation}
These local statements do not imply that the full field space is Euclidean.
If \((\R^n,g)\) were complete, the local isometry
\(\mathcal F:(\R^n,g)\to(\R^n,\delta)\) would be a covering map
\cite{doCarmo1992}.  Since the target is simply connected, the covering
would be one-to-one.  Any explicit collision therefore shows that the
pullback metric is geodesically incomplete.

The field equations retain the same simple local form.  Varying
eq.~\eqref{eq:general-action} gives
\begin{equation}
  (J_{\mathcal F})_{Ia}
  \left[-\nabla^2\mathcal Q^I
  +m^2(\mathcal Q^I-\mathcal Q_\star^I)\right]=0.
  \label{eq:general-eom-before-inverse}
\end{equation}
The Jacobian matrix is invertible pointwise, so
\begin{equation}
  (-\nabla^2+m^2)
  (\mathcal Q^I-\mathcal Q_\star^I)=0.
  \label{eq:general-free-eom}
\end{equation}
No inverse map for \(\mathcal F\) is used in this step; multiplication by the
pointwise matrix inverse \(J_{\mathcal F}^{-1}\) is sufficient.
Every finite critical point of the potential is a preimage of
\(\mathcal Q_\star\).  At such a point \(v_s\),
\begin{equation}
  \left.\partial_a\partial_bV\right|_{v_s}
  =f_\phi^2m^2g_{ab}(v_s).
  \label{eq:general-vacuum-hessian}
\end{equation}
The kinetic matrix is \(f_\phi^2g_{ab}(v_s)\), so all local normal modes have
mass \(m\).

For the Euclidean boundary condition
\(\mathcal Q-\mathcal Q_\star\to0\) at infinity, multiplying
eq.~\eqref{eq:general-free-eom} by
\(\mathcal Q-\mathcal Q_\star\) and integrating gives
\begin{equation}
  \int\dd^dX\,
  \left[
    \partial_\mu\mathcal Q^I\partial_\mu\mathcal Q^I
    +m^2(\mathcal Q^I-\mathcal Q_\star^I)^2
  \right]=0.
  \label{eq:general-no-saddle}
\end{equation}
Ordinary smooth finite-action solutions are therefore the constant vacua.
In particular, the different finite preimages are not connected by a smooth
finite-energy wall.  A transition between branches would have to probe an
incomplete end of field space, where an additional boundary prescription is
needed.

\subsection{Canonical variables and the global quantum test}

The local canonical transformation follows directly from the cotangent
lift.  For a field-space point \(\varphi\) with momentum \(\pi\), define
\begin{equation}
  \mathcal Q=\mathcal F(\varphi),
  \qquad
  \Pi=J_{\mathcal F}(\varphi)^{-\mathsf T}\pi.
  \label{eq:general-cotangent}
\end{equation}
Here \(J_{\mathcal F}^{-\mathsf T}\) is the inverse transpose of the Jacobian
matrix at \(\varphi\), not the derivative of a global inverse map.
Then
\begin{equation}
  \Pi_I\,\dd\mathcal Q^I
  =\pi_a\,\dd\varphi^a.
  \label{eq:general-one-form}
\end{equation}
The symplectic form is preserved on each local sheet.  Distinct phase-space
points nevertheless map to the same \((\mathcal Q,\Pi)\) whenever
\(\mathcal F\) has several preimages.

The corresponding Schr\"odinger operators act on
\(\cH_\varphi=L^2(\R^n,\dd^n\varphi)\).  On the common core
\(\cD=C_c^\infty(\R^n)\), set
\begin{equation}
  \widehat{\mathcal Q}_I=M_{\mathcal F_I},
  \qquad
  \widehat\Pi_I=-\ii\Delta_I,
  \qquad
  \Delta_I=(J_{\mathcal F}^{-\mathsf T})_{Ia}\partial_a.
  \label{eq:general-formal-operators}
\end{equation}
The unit determinant makes \(J_{\mathcal F}^{-1}\) a polynomial matrix.  With
the row convention
\((J_{\mathcal F})_{Ia}=\partial_a\mathcal F_I\), one has
\begin{equation}
  (J_{\mathcal F}^{-1})_{aI}
  =\frac{(\cof J_{\mathcal F})_{Ia}}
  {\det J_{\mathcal F}}
  =(\cof J_{\mathcal F})_{Ia}.
  \label{eq:general-cofactor}
\end{equation}
The Piola identity then gives
\begin{equation}
  \diver\Delta_I
  =\partial_a(\cof J_{\mathcal F})_{Ia}=0,
  \label{eq:general-piola}
\end{equation}
so \(-\ii\Delta_I\) is symmetric on \(\cD\)
\cite{KupfermanShachar2019}.  The inverse Jacobian also gives
\begin{equation}
  \Delta_I\mathcal F_J
  =(J_{\mathcal F}^{-1})_{aI}(J_{\mathcal F})_{Ja}
  =\delta_{IJ}.
  \label{eq:general-duality}
\end{equation}
It follows that \([\Delta_I,\Delta_J]\) annihilates every
\(\mathcal F_K\).  Since the one-forms \(\dd\mathcal F_K\) form a coframe,
the vector fields commute.  Hence
\begin{equation}
  [\widehat{\mathcal Q}_J,\widehat\Pi_I]
  =\ii\delta_{IJ},
  \qquad
  [\widehat\Pi_I,\widehat\Pi_J]=0
  \label{eq:general-formal-ccr}
\end{equation}
on \(\cD\).  These relations are formal.  They do not imply essential
self-adjointness, complete momentum flows, or regular Weyl relations.

The global test uses the number of real preimages
\begin{equation}
  \mathcal N_{\mathcal F}(\mathcal Q)
  =\#\{\varphi\in\R^n:\mathcal F(\varphi)=\mathcal Q\}.
  \label{eq:general-multiplicity}
\end{equation}
At a regular target value, this is equivalently the number of local inverse
branches above that value.
The area formula gives, for integrable test functions,
\begin{equation}
  \int_{\R^n}r(\mathcal F(\varphi))\,\dd^n\varphi
  =\int_{\R^n}
  \mathcal N_{\mathcal F}(\mathcal Q)r(\mathcal Q)\,\dd^n\mathcal Q
  \label{eq:general-area}
\end{equation}
\cite{EvansGariepy2015}.  Correspondingly,
\begin{equation}
  L^2(\R^n_\varphi)
  \simeq
  \int_{\R^n_{\mathcal Q}}^\oplus
  \C^{\mathcal N_{\mathcal F}(\mathcal Q)}
  \,\dd^n\mathcal Q.
  \label{eq:general-direct-integral}
\end{equation}
The usual coordinate tuple has joint spectral multiplicity one.  If
\(\mathcal N_{\mathcal F}\) equals three on one open set and one on another,
the tuple \(M_{\mathcal F_I}\) cannot be related to the usual coordinates by
a unitary operator.  A regular representation of the finite-dimensional
Weyl relations also has constant spectral multiplicity, as follows from the
Stone--von Neumann classification \cite{Folland1989}.  Nonconstant
\(\mathcal N_{\mathcal F}\) therefore obstructs regular exponentiation even
though eq.~\eqref{eq:general-formal-ccr} holds locally.

\section{Three scalar fields from the Alp\"oge map}
\label{sec:alpoge-three-scalar}

\subsection{The map and our conventions}

The input from ref.~\cite{Alpoge2026Map} is a polynomial map, not a
three-dimensional spacetime model.  We use it as a map of field space and
place the resulting scalar fields on a \(d\)-dimensional Euclidean
background.  This distinction is summarized in
table~\ref{tab:alpoge-dictionary}.

\begin{table}[t]
\centering
\small
\begin{tabular}{p{0.25\textwidth}p{0.30\textwidth}p{0.34\textwidth}}
\hline
Alp\"oge announcement & Present paper & Interpretation\\
\hline
\((x,y,z)\in\R^3\) &
\(\varphi^a(X)=(x(X),y(X),z(X))\) &
Dimensionless field-space coordinates promoted to three real scalar fields\\
\((A,B,C)\) &
\(\mathcal F_3=(-A/2,B,C)\) &
Linear output normalization used to set the Jacobian to \(+1\)\\
\(\det J_{(A,B,C)}=-2\) &
\(\det J_{\mathcal F_3}=1\) &
The local functional measure has unit density\\
\((-1/4,0,0)\) &
\(\mathcal Q_\star^{(3)}=(1/8,0,0)\) &
The target around which the massive potential is centered\\
Weighted algebraic scaling &
Fixed coefficients in the scalar action &
The scaling is not a continuous symmetry of the massive shifted theory\\
\hline
\end{tabular}
\caption{Dictionary between the announced algebraic map and the
three-scalar model used here.}
\label{tab:alpoge-dictionary}
\end{table}

Write \(s=1+xy\).  The announced map
\(\mathcal L=(A,B,C)\) is
\begin{align}
  A&=s^3z+y^2s(4+3xy),
  \label{eq:alpoge-A}\\
  B&=y+3xs^2z+3xy^2(4+3xy),
  \label{eq:alpoge-B}\\
  C&=2x-3x^2y-x^3z.
  \label{eq:alpoge-C}
\end{align}
Its Jacobian is \(-2\).  We normalize the first target coordinate and use
\begin{equation}
  \mathcal F_3(x,y,z)
  =\left(-\frac{A}{2},B,C\right),
  \qquad
  \det J_{\mathcal F_3}=1.
  \label{eq:alpoge-normalized-map}
\end{equation}
The three points
\begin{equation}
  v_0=\left(0,0,-\frac14\right),
  \qquad
  v_+=\left(1,-\frac32,\frac{13}{2}\right),
  \qquad
  v_-=\left(-1,\frac32,\frac{13}{2}\right)
  \label{eq:alpoge-three-vacua}
\end{equation}
all map to
\begin{equation}
  \mathcal Q_\star^{(3)}
  =\left(\frac18,0,0\right).
  \label{eq:alpoge-target-star}
\end{equation}
Since the Jacobian is nonzero at all three points, the inverse function
theorem gives disjoint source neighborhoods
\(\mathcal U_s\ni v_s\) and a common target neighborhood
\(\mathcal V_\star^{(3)}\ni\mathcal Q_\star^{(3)}\) such that each restriction
\(\mathcal F_3|_{\mathcal U_s}\) is a diffeomorphism onto
\(\mathcal V_\star^{(3)}\).  We denote its local inverse by
\(\mathcal G_s=(\mathcal F_3|_{\mathcal U_s})^{-1}\).  These three maps are
the local inverse branches.  They cannot be combined into a global
single-valued inverse because
\(\mathcal G_s(\mathcal Q_\star^{(3)})=v_s\) gives three different values.

The fiber equation is controlled by a cubic.  For the unnormalized target
\((A,B,C)\), introduce
\begin{equation}
  P_{A,B,C}(T)=CT^3-2T^2+BT-2A.
  \label{eq:alpoge-fiber-polynomial}
\end{equation}
On the chart \(x\neq0\), \(t=y+1/x\) satisfies
\begin{equation}
  P_{A,B,C}(t)=0,
  \qquad
  P'_{A,B,C}(t)=\frac2x.
  \label{eq:alpoge-fiber-identities}
\end{equation}
To see the reconstruction directly, substitution gives
\begin{equation}
  B=4t+\frac2x-3Ct^2,
  \qquad
  2A=Ct^3-2t^2+Bt.
  \label{eq:alpoge-coordinate-identities}
\end{equation}
For a simple root \(\tau\), let
\(\rho=P'_{A,B,C}(\tau)\).  The corresponding source point is
\begin{equation}
  x=\frac2\rho,
  \qquad
  y=\tau-\frac\rho2,
  \qquad
  z=\frac54\rho^2-\frac32\tau\rho-\frac C8\rho^3.
  \label{eq:alpoge-reconstruction}
\end{equation}
Thus a generic target has three complex preimages
\cite{DorkyCubic2026}.

The vacuum target requires one small qualification.  Here \(C=0\), so the
leading cubic coefficient vanishes and the \(x=0\) chart must be retained.
On that chart, \(B=C=0\) implies \(y=0\), while \(A=-1/4\) fixes
\(z=-1/4\), giving \(v_0\).  On the \(x\neq0\) chart the remaining fiber
equation is
\begin{equation}
  -2T^2+\frac12=0.
  \label{eq:alpoge-star-degenerate-cubic}
\end{equation}
Its two roots reconstruct \(v_\pm\).  This verifies that the three points in
eq.~\eqref{eq:alpoge-three-vacua} exhaust the finite real fiber over
\(\mathcal Q_\star^{(3)}\).

For the quantum argument we also need a target with one real preimage.
Choose the unnormalized value \((A,B,C)=(1,3,1)\), corresponding to
\begin{equation}
  \mathcal Q_0^{(3)}
  =\left(-\frac12,3,1\right).
  \label{eq:alpoge-target-zero}
\end{equation}
The fiber polynomial becomes
\begin{equation}
  P_0^{(3)}(T)
  =T^3-2T^2+3T-2
  =(T-1)(T^2-T+2).
  \label{eq:alpoge-one-root-fiber}
\end{equation}
It has one real root and the exact preimage
\begin{equation}
  \mathcal F_3(1,0,1)
  =\mathcal Q_0^{(3)}.
  \label{eq:alpoge-one-root-point}
\end{equation}
Indeed, \(T=1\) has \(P_0^{(3)\prime}(1)=2\), and
eq.~\eqref{eq:alpoge-reconstruction} immediately gives
\((x,y,z)=(1,0,1)\).
The nonreal pair stays nonreal under a small target perturbation, so a
neighborhood \(\mathcal V_0^{(3)}\) has one real sheet.

\subsection{The three-scalar action}

Substituting eqs.~\eqref{eq:alpoge-A}--\eqref{eq:alpoge-C} into the general
construction gives the complete interacting action
\begin{equation}
\begin{split}
  S_3
  =\frac{f_\phi^2}{2}\int\dd^dX\,
  \bigg\{&
    \frac14(\partial_\mu A)^2
    +(\partial_\mu B)^2
    +(\partial_\mu C)^2\\
    &+m^2\left[
      \left(-\frac A2-\frac18\right)^2+B^2+C^2
    \right]\bigg\}.
  \label{eq:alpoge-three-scalar-action}
\end{split}
\end{equation}
The derivative interactions come from the three polynomial composites
\((A,B,C)\).  The potential contains the corresponding nonderivative
interactions.  Their relative coefficients are fixed by the unit-Jacobian
map; the independent physical parameters of this minimal model are
\(f_\phi\) and \(m\).

The original fields do not represent three decoupled particle species.
They are nonlinear coordinates on one flat target field space.  Algebraically,
\(xy\) controls the shape variable \(s\), \(x\) sets the scale that appears in
the reconstruction relation \(P'(t)=2/x\), and \(z\) supplies the direction
that is linear in the map.  The propagating normal modes near a vacuum are
the three combinations
\begin{equation}
  q^I=f_\phi\left[
  \mathcal F_3^I(\varphi)-\mathcal Q_\star^{(3)I}\right],
  \label{eq:alpoge-normal-modes}
\end{equation}
not \(x,y,z\) separately.

The algebraic map is covariant under
\begin{equation}
  (x,y,z)\longrightarrow
  (\lambda x,\lambda^{-1}y,\lambda^{-2}z),
  \qquad
  (A,B,C)\longrightarrow
  (\lambda^{-2}A,\lambda^{-1}B,\lambda C).
  \label{eq:alpoge-weighted-scaling}
\end{equation}
The Euclidean target metric and the shifted massive potential break this
continuous scaling.  The \(\lambda=-1\) transformation remains an exact
\(\mathbb Z_2\) symmetry:
\begin{equation}
  (x,y,z)\longrightarrow(-x,-y,z),
  \qquad
  (A,B,C)\longrightarrow(A,-B,-C).
  \label{eq:alpoge-z2}
\end{equation}
It fixes \(v_0\) and exchanges \(v_+\) with \(v_-\).

The action has exactly the three vacua in
eq.~\eqref{eq:alpoge-three-vacua}.  At \(v_0\), for example,
\begin{equation}
  J_{\mathcal F_3}(v_0)
  =
  \begin{pmatrix}
    0&0&-1/2\\
    -3/4&1&0\\
    2&0&0
  \end{pmatrix},
  \qquad
  g^{(3)}(v_0)
  =
  \begin{pmatrix}
    73/16&-3/4&0\\
    -3/4&1&0\\
    0&0&1/4
  \end{pmatrix},
  \label{eq:alpoge-vacuum-metric}
\end{equation}
and \(\det g^{(3)}(v_0)=1\).  Equation
\eqref{eq:general-vacuum-hessian} then gives three modes of mass \(m\).
The same conclusion holds at \(v_\pm\), despite the less diagonal coordinate
matrices there.

The exact equations of motion are
\begin{equation}
  (-\partial^2+m^2)
  \left[\mathcal F_3^I(\varphi)
  -\mathcal Q_\star^{(3)I}\right]=0.
  \label{eq:alpoge-eom}
\end{equation}
Small Lorentzian excitations are therefore ordinary massive waves on each
local sheet.  With Euclidean finite-action boundary conditions, the only
smooth solutions are the three constant vacua.  The model has no smooth
finite-energy wall connecting them at finite field values.

\subsection{Global quantum structure}

The three vacua have the same local perturbative physics, but they do not
collapse to one global Schr\"odinger representation.  On
\(\mathcal V_\star^{(3)}\), the position tuple
\(M_{\mathcal F_3^I}\) has joint spectral multiplicity three.  On
\(\mathcal V_0^{(3)}\), it has multiplicity one.  Therefore no unitary
operator on \(L^2(\R^3)\) can satisfy
\begin{equation}
  U M_{\varphi^I}U^{-1}=M_{\mathcal F_3^I(\varphi)}
  \qquad (I=1,2,3).
  \label{eq:alpoge-no-unitary}
\end{equation}
The same variation in multiplicity rules out strongly continuous momentum
groups satisfying the Weyl relations with these position operators.

This statement does not modify the branchwise equivalence theorem.  On each
sheet, the formal momenta
\begin{equation}
  \widehat\Pi_I
  =-\ii(J_{\mathcal F_3}^{-\mathsf T})_{Ia}\partial_a
  \label{eq:alpoge-formal-momenta}
\end{equation}
are symmetric on \(C_c^\infty(\R^3)\) and satisfy the canonical commutators.
The obstruction appears only when one asks for a single self-adjoint,
regular, global realization that covers both the one-sheeted and
three-sheeted target regions.

\section{A four-scalar model}
\label{sec:four-scalar}

\subsection{The polynomial map and its fibers}

We now promote the constant parameter in the Jacobian-neutral rational frame
of ref.~\cite{DorkyFamily2026} to a fourth scalar \(w\), producing a map
whose generic fiber is quintic.  An independent weighted-lift construction
realizing every generic fiber degree at least three was given in
ref.~\cite{Gallagher2026}.  Define
\begin{equation}
  \sigma=1+xy,
  \qquad
  u=\sigma(1-x^2z),
  \qquad
  \eta=y-xz-x^2yz,
  \label{eq:four-sigma-u-eta}
\end{equation}
so that \(u=1+x\eta\), and set
\begin{align}
  a={}&\sigma(y^2-\sigma^2z)
  +\frac12\sigma^2\eta^2(9+w-4wu),
  \label{eq:four-a}\\
  b={}&-2y+x\sigma\eta^2(12+2w-5wu),
  \label{eq:four-b}\\
  c={}&x-x^3z.
  \label{eq:four-c}
\end{align}
The unnormalized and normalized maps are
\begin{equation}
  G_4=(a,b,c,w),
  \qquad
  \mathcal F_4=\left(-\frac a2,b,c,w\right).
  \label{eq:four-map}
\end{equation}
A direct calculation gives
\begin{equation}
  \det J_{G_4}=-2,
  \qquad
  \det J_{\mathcal F_4}=1.
  \label{eq:four-det}
\end{equation}

The constant determinant is transparent in a rational frame.  On \(x\neq0\)
introduce
\begin{equation}
  t=y+\frac1x,
  \qquad
  r=\frac2x,
  \qquad
  g_W(U)=-3U^2+8U-5+W(U-1)^3,
  \label{eq:four-rational-frame}
\end{equation}
and \(h(T,C,W)=T^2g_W(CT)\).  Substitution gives
\begin{equation}
  b=r-\partial_T h(t,c,w),
  \qquad
  2a=h(t,c,w)+tb.
  \label{eq:four-rational-identities}
\end{equation}
The two Jacobian factors are \(r/2\) and \(-2x\), whose product is
\(-2\).  Since the final expressions are polynomial, the identity extends
through \(x=0\).

For a target \((A,B,C,W)\) of \(G_4\), the fiber polynomial is
\begin{equation}
  P_{A,B,C,W}(T)=h(T,C,W)+BT-2A.
  \label{eq:four-fiber-polynomial}
\end{equation}
At a preimage,
\begin{equation}
  P(t)=0,
  \qquad
  P'(t)=\frac2x.
  \label{eq:four-fiber-identities}
\end{equation}
A simple root \(\tau\), with \(\rho=P'(\tau)\), reconstructs
\begin{equation}
  x=\frac2\rho,
  \qquad
  y=\tau-\frac\rho2,
  \qquad
  z=\frac{x-C}{x^3},
  \qquad
  w=W.
  \label{eq:four-reconstruction}
\end{equation}
For \(CW\neq0\), the fiber equation is generically of degree five.

The normalized target
\begin{equation}
  \mathcal Q_\star^{(4)}=(-4,16,1,-3)
  \label{eq:four-target-star}
\end{equation}
corresponds to \((A,B,C,W)=(8,16,1,-3)\).  Its fiber polynomial factors as
\begin{equation}
\begin{split}
  P_\star^{(4)}(T)
  &=-3T^5+6T^4-T^3-2T^2+16T-16\\
  &=-(T-1)(T-2)(3T^3+3T^2+4T+8).
  \label{eq:four-factor}
\end{split}
\end{equation}
The cubic factor is strictly increasing because its derivative is
\(9T^2+6T+4>0\).  The fiber therefore contains three real points and one
nonreal conjugate pair.  Two real preimages are
\begin{equation}
  q_1=\left(\frac19,-8,-648,-3\right),
  \qquad
  q_2=\left(-\frac1{26},28,18252,-3\right).
  \label{eq:four-rational-preimages}
\end{equation}
These points follow from
\begin{equation}
  {P_\star^{(4)}}'(1)=18,
  \qquad
  {P_\star^{(4)}}'(2)=-52.
  \label{eq:four-rational-root-derivatives}
\end{equation}
The third real root is the unique solution
\(\alpha\simeq-1.403615886831572\) of
\begin{equation}
  3\alpha^3+3\alpha^2+4\alpha+8=0.
  \label{eq:four-alpha}
\end{equation}
Writing \(x_3=2/{P_\star^{(4)}}'(\alpha)\), its preimage is
\begin{equation}
  q_3=\left(
  x_3,\,
  \alpha-\frac{{P_\star^{(4)}}'(\alpha)}2,\,
  \frac{x_3-1}{x_3^3},\,
  -3\right).
  \label{eq:four-third-preimage}
\end{equation}
All five roots are simple, so the three real preimages extend to a common
three-sheeted target neighborhood \(\mathcal V_\star^{(4)}\).

The comparison target
\begin{equation}
  \mathcal Q_0^{(4)}=(0,6,1,1)
  \label{eq:four-target-zero}
\end{equation}
has the unique real preimage
\begin{equation}
  q_0=\left(\frac13,-3,-18,1\right).
  \label{eq:four-one-preimage}
\end{equation}
For this target the fiber polynomial is
\begin{equation}
  P_0^{(4)}(T)
  =T^5-6T^4+11T^3-6T^2+6T.
  \label{eq:four-zero-polynomial}
\end{equation}
Its derivative obeys
\begin{equation}
\begin{split}
  {P_0^{(4)}}'(T)
  &=5T^4-24T^3+33T^2-12T+6\\
  &=5\left(T^2-\frac{12}{5}T+\frac3{10}\right)^2
  +\frac65(T-2)^2+\frac34>0.
  \label{eq:four-one-root-positive}
\end{split}
\end{equation}
Strict positivity persists under a small target perturbation, giving an open
one-sheeted neighborhood \(\mathcal V_0^{(4)}\).

\subsection{Four fields and the extra coordinate}

The four-scalar action is the \(n=4\) case of
eq.~\eqref{eq:general-action},
\begin{equation}
\begin{split}
  S_4
  =\frac{f_\phi^2}{2}\int\dd^dX\,
  \big[
    (J_{\mathcal F_4}^{\mathsf T}J_{\mathcal F_4})_{ab}
    \partial_\mu\phi^a\partial_\mu\phi^b
    +m^2\lvert\mathcal F_4(\phi)
    -\mathcal Q_\star^{(4)}\rvert^2
  \big],
  \label{eq:four-action}
\end{split}
\end{equation}
where \(\phi=(x,y,z,w)\).  The fourth output is \(w\), but \(w\) also
appears in \(a\) and \(b\).  It is therefore not a decoupled spectator.
Changing its target value changes \(g_W\) and hence the quintic fiber
polynomial seen by the other three fields.

The potential has exactly three real vacua over
\(\mathcal Q_\star^{(4)}\).  Near each one, the four combinations
\begin{equation}
  q^I=f_\phi\left[
  \mathcal F_4^I(\phi)-\mathcal Q_\star^{(4)I}\right]
  \label{eq:four-normal-modes}
\end{equation}
are free fields of mass \(m\).  The exact equations are
\begin{equation}
  (-\partial^2+m^2)
  \left[\mathcal F_4^I(\phi)
  -\mathcal Q_\star^{(4)I}\right]=0.
  \label{eq:four-eom}
\end{equation}
The field-space metric is flat with unit determinant but incomplete, as the
three-point fiber makes global completeness impossible.

The quantum result follows without a new operator calculation.  The
position tuple has multiplicity three on
\(\mathcal V_\star^{(4)}\) and one on
\(\mathcal V_0^{(4)}\).  It is not unitarily equivalent to the ordinary
four-coordinate Schr\"odinger tuple, and the formal momenta cannot generate
a regular Weyl representation with these positions.  The fourth field
changes the algebraic family but not the spectral mechanism.

The invariant hyperplane \(w=-3\) gives a separate three-scalar map of
generic degree five.  It retains useful exact information about the
four-field family, including the same three real vacua, and coincides with
the explicit generic-degree-five member of the three-variable family in
ref.~\cite{DorkyFamily2026}.  Because it is not the minimal three-scalar
example, we keep the calculation in appendix~\ref{app:degree-five-slice}.

\section{Comparing the three- and four-scalar models}
\label{sec:comparison}

The two examples are summarized in table~\ref{tab:model-comparison}.  Their
polynomial degrees and target values differ, but their physical
interpretation is the same.

\begin{table}[t]
\centering
\small
\begin{tabular}{c c c c}
\hline
Model & Generic complex fiber & Three-sheet target & One-sheet target\\
\hline
\(\mathcal F_3\) &
\(3\) &
\((1/8,0,0)\) &
\((-1/2,3,1)\)\\
\(\mathcal F_4\) &
\(5\) &
\((-4,16,1,-3)\) &
\((0,6,1,1)\)\\
\hline
\end{tabular}
\caption{Algebraic data used to diagnose the global quantum structure.  The
sheet numbers in the last two columns count real preimages, equivalently
local inverse branches, on small open neighborhoods of the displayed targets.}
\label{tab:model-comparison}
\end{table}

Three ingredients are common.  First, \(\det J_{\mathcal F}=1\) removes any
local measure density and makes the formal momenta symmetric through the
Piola identity.  Second, the map is nonproper.  Real branches can escape to
field-space infinity while their images remain finite, so the number of
real preimages can change without a critical point.  Third, the
pullback metric is flat but incomplete.  The same incomplete end appears
geometrically in the sigma model and spectrally in the failure of complete
momentum flows.

The area formula also explains the branch factor in the functional
integral.  If a connected spacetime configuration is restricted to a common
\(k\)-sheeted target neighborhood, continuity makes the local-sheet label
constant on spacetime.  The smooth branch-restricted path integral is then a
sum of \(k\) identical free sectors, not a local factor independently chosen
at every spacetime point.  A global path integral must additionally specify
what happens when fields approach the incomplete ends.

The pullback operator makes the surviving global relation explicit:
\begin{equation}
  C_{\mathcal F}:L^2(\R^n_{\mathcal Q})
  \longrightarrow L^2(\R^n_\varphi),
  \qquad
  (C_{\mathcal F}\psi)(\varphi)
  =\psi(\mathcal F(\varphi)).
  \label{eq:comparison-pullback}
\end{equation}
The area formula gives
\begin{equation}
  C_{\mathcal F}^\dagger C_{\mathcal F}
  =M_{\mathcal N_{\mathcal F}}.
  \label{eq:comparison-pullback-norm}
\end{equation}
On a three-sheeted region its range is the diagonal subspace
\((\psi,\psi,\psi)\) of the multiplicity-three spectral fiber.  The operator
is an intertwiner, but it is neither onto nor unitary.

The additional \(w\) field in \(\mathcal F_4\) changes the fiber polynomial
from cubic to quintic and promotes a coefficient of the three-variable
family to a dynamical field.  It does not remove the branchwise free
description.  This comparison indicates that the quantum obstruction is
controlled by preimage multiplicity rather than by the number of
fields or the degree of a particular polynomial presentation.

\section{Discussion and outlook}
\label{sec:discussion}

The three-scalar model provides the shortest physical realization of the
global issue.  Its local Lagrangian looks highly interacting in the
\((x,y,z)\) coordinates, yet every perturbative sector is exactly a free
massive triplet.  The nontrivial information lies in the way the local
coordinate patches fit together.  Three finite vacua are three preimages of
the same target configuration, each defining a local inverse branch, and the
preimage count changes elsewhere in target space.

This does not conflict with the equivalence theorem.  That theorem applies
after one chooses an invertible local redefinition and compatible
asymptotic data.  It does not identify several distinct local inverse
branches with a single field coordinate.  Likewise, the formal canonical
commutators are correct on their test-function domain.  Their global
exponentiation fails because a regular Weyl representation cannot carry a
position tuple whose spectral multiplicity changes from one open set to
another.

The four-scalar model shows that the same physics survives when the fiber
equation is quintic and one coefficient becomes a field.  The \(w=-3\) slice
in appendix~\ref{app:degree-five-slice} provides a useful intermediate
example, but the Alp\"oge map remains the simpler starting point for the
three-scalar theory.

Several questions require additional input rather than further local
algebra.  A nonperturbative path integral must choose boundary conditions at
the incomplete ends of field space.  Observables may be restricted to be
branch-blind, or the local-sheet label may be treated as additional global
data.  Interacting target potentials \(V(\mathcal Q)\) can also be pulled
back, preserving the same local geometry while changing the spectrum and
classical solutions.  These choices determine a quantum completion; the
unit Jacobian alone does not.

\appendix

\section{A generic-degree-five three-scalar slice}
\label{app:degree-five-slice}

This appendix retains the former three-scalar specialization of the
four-field map.  It is the invariant slice \(w=-3\), and the resulting
generic-degree-five map is the explicit degree-five member of the public
construction in ref.~\cite{DorkyFamily2026}.  The example is useful because
its three real vacua coincide with the first three coordinates of the
four-field vacua and because all relevant spectral statements can be checked
within three variables.

Define
\begin{equation}
  f_5:\R^3\longrightarrow\R^3,
  \qquad
  f_5(x,y,z)=\left(-\frac{a_5}{2},b_5,c_5\right),
  \label{eq:five-map}
\end{equation}
where \(\sigma,u,\eta\) are given in
eq.~\eqref{eq:four-sigma-u-eta} and
\begin{align}
  a_5&=\sigma(y^2-\sigma^2z)+3\sigma^2\eta^2(1+2u),
  \label{eq:five-a}\\
  b_5&=-2y+3x\sigma\eta^2(2+5u),
  \label{eq:five-b}\\
  c_5&=x-x^3z.
  \label{eq:five-c}
\end{align}
The block form of \(J_{\mathcal F_4}\) at fixed \(w\) gives
\begin{equation}
  \det J_{f_5}=1.
  \label{eq:five-det}
\end{equation}

For the target
\begin{equation}
  q_\star^{(5)}=(-4,16,1),
  \label{eq:five-target-star}
\end{equation}
the fiber polynomial is \(P_\star^{(4)}(T)\) in
eq.~\eqref{eq:four-factor}.  Its three real preimages are
\begin{align}
  v_1^{(5)}&=\left(\frac19,-8,-648\right),
  &
  v_2^{(5)}&=\left(-\frac1{26},28,18252\right),
  \label{eq:five-rational-vacua}\\
  v_3^{(5)}&=\left(x_\alpha,
  \alpha-\frac{\rho_\alpha}{2},
  \frac{x_\alpha-1}{x_\alpha^3}\right),
  &
  x_\alpha&=\frac2{\rho_\alpha},
  \qquad
  \rho_\alpha={P_\star^{(4)}}'(\alpha).
  \label{eq:five-algebraic-vacuum}
\end{align}
Numerically,
\begin{equation}
  v_3^{(5)}
  \simeq
  (-0.0183679739102,\,53.0389701539,\,164331.557504).
  \label{eq:five-third-vacuum-numeric}
\end{equation}

The associated scalar model is
\begin{equation}
  S_{3,5}
  =\frac{f_\phi^2}{2}\int\dd^dX
  \left[
    (J_{f_5}^{\mathsf T}J_{f_5})_{ab}
    \partial_\mu\varphi^a\partial_\mu\varphi^b
    +m^2\lvert f_5(\varphi)-q_\star^{(5)}\rvert^2
  \right].
  \label{eq:five-action}
\end{equation}
Its metric is positive and flat with unit determinant.  The three points in
eqs.~\eqref{eq:five-rational-vacua} and
\eqref{eq:five-algebraic-vacuum} are its three vacua, and all local modes
have mass \(m\).

At the first rational vacuum,
\begin{equation}
  J_{f_5}(v_1^{(5)})
  =
  \begin{pmatrix}
    352&-4&1/1458\\
    0&-2&0\\
    25&0&-1/729
  \end{pmatrix},
  \label{eq:five-vacuum-jacobian}
\end{equation}
so
\begin{equation}
  g^{(5)}(v_1^{(5)})
  =
  \begin{pmatrix}
    124529&-1408&151/729\\
    -1408&20&-2/729\\
    151/729&-2/729&5/2125764
  \end{pmatrix},
  \qquad
  \det g^{(5)}(v_1^{(5)})=1.
  \label{eq:five-vacuum-metric}
\end{equation}
The anisotropic coordinate entries do not affect the physical masses,
because \(f_5-q_\star^{(5)}\) supplies local normal coordinates.

The real spectral multiplicity also varies in this slice.  Take
\begin{equation}
  q_0^{(5)}=(0,13,-1).
  \label{eq:five-target-zero}
\end{equation}
The corresponding fiber polynomial and derivative are
\begin{align}
  P_0^{(5)}(T)&=3T^5+6T^4+T^3-2T^2+13T,
  \label{eq:five-one-root-polynomial}\\
  {P_0^{(5)}}'(T)&=15T^4+24T^3+3T^2-4T+13.
  \label{eq:five-one-root-derivative}
\end{align}
Using
\(24T^3\geq-12T^4-12T^2\) and
\(4\lvert T\rvert\leq T^2+4\), one finds
\begin{equation}
  {P_0^{(5)}}'(T)
  \geq3T^4-10T^2+9
  =3\left(T^2-\frac53\right)^2+\frac23>0.
  \label{eq:five-one-root-positive}
\end{equation}
The unique real root \(T=0\) gives
\begin{equation}
  f_5\left(\frac2{13},-\frac{13}{2},\frac{2535}{8}\right)
  =q_0^{(5)}.
  \label{eq:five-one-preimage}
\end{equation}
Consequently the same multiplicity-three versus multiplicity-one quantum
obstruction occurs for this degree-five slice.

\acknowledgments

OpenAI Codex was used to assist with algebraic exploration, literature
searches, drafting, and typesetting.  BZ is supported by the Fundamental
Research Funds for the Central Universities (010-63263123).

\bibliographystyle{JHEP}
\bibliography{references}

\end{document}